# Generative Models Improve Radiomics Performance in Different Tasks and Different Datasets: An Experimental Study

**Short Title**: Generative Model Improve Radiomics Performance

Junhua Chen, MS [a,1]; Inigo Bermejo, PhD [1]; Andre Dekker, PhD [1]; Leonard Wee, PhD [1]

[1]. Department of Radiation Oncology (MAASTRO), GROW School for Oncology and Developmental Biology, Maastricht University Medical Centre+, Maastricht, 6229 ET, Netherlands

[a]. Corresponding author, Tel.: +31 0684 6149 49 E-mail address: junhua.chen@maasrto.nl

**Address of Corresponding author:** Department of Radiation Oncology (MAASTRO), GROW School for Oncology and Developmental Biology, Maastricht University Medical Centre+, Maastricht, 6229 ET, Netherlands

**Abstract:**

**Background:** Radiomics is an active area of research focusing on high throughput feature extraction from medical images with a wide array of applications in clinical practice, such as clinical decision support in oncology. However, noise in low dose computed tomography (CT) scans can impair the accurate extraction of radiomic features. In this article, we investigate the possibility of using deep learning generative models to improve the performance of radiomics from low dose CTs.

**Methods:** We used two datasets of low dose CT scans – NSCLC Radiogenomics and LIDC-IDRI – as test datasets for two tasks – pre-treatment survival prediction and lung cancer diagnosis. We used encoder-decoder networks and conditional generative adversarial networks (CGANs) trained in a previous study as generative models to transform low dose CT images into full dose CT images. Radiomic features extracted from the original and improved CT scans were used to build two classifiers – a support vector machine (SVM) and a deep attention based multiple instance learning model – for survival prediction and lung cancer diagnosis respectively. Finally, we compared the performance of the models derived from the original and improved CT scans.

**Results:** Encoder-decoder networks and CGANs improved the area under the curve (AUC) of survival prediction from 0.52 to 0.57 (p-value<0.01). On the other hand, Encoder-decoder network and CGAN can improve the AUC of lung cancer diagnosis from 0.84 to 0.88 and 0.89 respectively (p-value<0.01). Moreover, there are no statistically significant differences in improving AUC by using encoder-decoder network and CGAN (p-value=0.34) when networks trained at 75 and 100 epochs.

**Conclusion:** Generative models can improve the performance of low dose CT-based radiomics in different tasks. Hence, denoising using generative models seems to be a necessary pre-processing step for calculating radiomic features from low dose CTs.

# 1. Introduction

In recent years, as an active research topic, radiomics [1] has been applied to clinical-decision support in oncology in a range of cancers (lung cancers, [2] head and neck cancer, [3] rectal cancer [4]) and multiple medical imaging modalities (computed tomography (CT), [3] magnetic resonance imaging (MRI), [5] and positron emission tomography (PET)). [2]

Following the ALARA (As Low As Reasonably Achievable) principle [6], low dose CTs has become popular as the preferred imaging method for screening and monitoring populations at risk [7]. As a tradeoff of low radiation exposure, low dose CTs' image quality is inferior to that of full dose CTs', due to the higher noise levels present in low dose CTs. Image quality and noise impact the repeatability and reproducibility of radiomic features [8] as well as their robustness [9]. In other words, radiomic features extracted from low dose CTs have lower reliability than the counterparts extracted from full dose CTs. Therefore, prediction models or computer aided diagnosis systems based on radiomic features from low dose CTs will likely be less robust and accurate than those based on radiomic features from full dose CT. Improving the performance of radiomics calculated from low dose CT in different tasks and datasets is therefore a timely and potentially impactful research topic.

One approach is to denoise low dose CT scans [10] and to recalculate the radiomic features based on the denoised CT. The aim of this article, is to answer the question: should we regard denoising as a preprocessing step for radiomic feature extraction from low dose CT? Image denoising can be regarded as a special case of domain adaptation [11], from low dose CT images to full dose style CT images [12]. Many methods have been proposed to perform this transformation [13][14], but recently deep learning [15] based generative models have garnered special attention and achieved state-of-art results [16][17][18][19]. We will use generative models to denoise low dose CT scans and improve the reliability of radiomic features.

In addition, we will explore whether more reliable radiomic features result in models with better performance using two real applications of radiomics: pre-treatment survival prediction [1] and cancer diagnosis [22][23]. The cancer diagnosis task will be based on [24], in which lung cancer diagnosis was approached as a multiple instance learning (MIL) problem [25] where nodules in each CT scan were regarded as instances. The authors used radiomic features as the input and deep attention based MIL [26] as the MIL problem solver for the sake of interpretability. The authors reported a mean precision of 0.807 with a standard error of the mean (SEM) of 0.069, a recall of 0.870 (SEM 0.061), and an area under the curve (AUC) of 0.842 (SEM 0.074) by using this method.

The most related literature to this article is [27], where the authors trained three generative models – encoder-decoder networks [17], conditional generative adversarial networks (GANs) [28] and cycle GANs [29] – using full dose CTs and simulated paired high-noise low dose CTs. Finally, they showed that radiomic features extracted from low dose CT scans (low-noise CT and high-noise CT) denoised by the models had improved reproducibility. The main differences between [27] and this article is that: 1) we use pre-trained generative models; 2) we use real (not simulated ) low dose CTs; and 3) we focus on the improvement in radiomics-based model performance instead of feature reproducibility.

To the authors' best knowledge, this is the first effort to improve the performance of radiomics-based models from features extracted from low dose CT scans. Source code, Radiomics features, data for statistical analysis and supplementary materials of this article are available online at https://gitlab.com/UM-CDS/low-dose-ct-denoising/-/tree/Experimental_Study.

## 2. Methods

Institutional Review Board approval was not applicable for this study, since the primary source of data was an open access collection on The Cancer Imaging Archive (National Institutes of Health) [30] and all patients' personal information had been removed from CT scans. This dataset has been used for this study in accordance with the Creative Commons Attribution-NonCommercial 3.0 Unported (CC BY-NC) conditions. The flowchart in Figure 1 summarizes our study methodology.

*2.1 Model Development*

Based on [27], we selected two generative models – encoder-decoder networks and CGANs – that achieved good performance in improving radiomics reproducibility as the experimental models for this study. Moreover, we took the same architecture of encoder-decoder network and CGANs presented in [27].

Training of encoder-decoder networks and CGANs requires paired low dose and full dose versions of the same CT scan. Although there is an open access dataset containing this kind of scans [31], the exposure of low dose CT scans in the dataset is higher - 50 milliampere-seconds (mAs) - than in many low dose CT scanning situations. For example, CT scans in the non-small cell lung cancer (NSCLC) Radiogenomics dataset were scanned from 1 to 400 mAs [32] and over half CT images scanned with an exposure lower or equal to 5 mAs. Models trained from the dataset described in [31] may be have a bad performance in much lower CT scans. The noise power of high noise images (used to train the models) in [27] is 25 times than that in [31]. For this reason, we used trained models from paper [27] without re-training to denoise low dose CT images. The source code and pre-trained models can be found at https://gitlab.com/UM-CDS/low-dose-ct-denoising/.

*2.2 Data Acquisition*

As mentioned in the Introduction, we will apply pretrained generative models to improve the performance of low CT radiomics-based models in two tasks: pre-treatment survival prediction and lung cancer diagnosis. For this purpose, we chose the NSCLC Radiogenomics dataset [32] for survival prediction and the Lung Image Database Consortium image collection (LIDC-IDRI) for lung cancer diagnosis [33], because they contain the necessary mask of the region of interest (ROI) for calculating the radiomics features and the images were scanned with low radiation exposure.

NSCLC Radiogenomics is a unique radiogenomic dataset from a cohort of 211 patients with NSCLC [34], from which we used low dose CT images, their respective segmentation masks and clinical data for survival prediction. The LIDC-IDRI dataset contains 1018 clinical chest CT scans, along with 157 patients' diagnoses. We used the diagnoses and their respective CT scans for the lung cancer diagnosis task. Finally, 106 samples of the NSCLC Radiogenomics were selected for survival prediction and 110 samples from LIDC-IDRI for lung cancer diagnosis. The index of selected samples for further investigation can be found in Supplementary Tables 1 and 2. The average radiation exposure of selected samples was $38.65 \pm 81.97$ mAs (±=SEM) in NSCLC Radiogenomics and $145.79 \pm 174.57$ mAs in LIDC-IDRI. The distributions of radiation exposure for the two datasets are shown in Supplementary Figure 1.

*2.3 Extraction of Radiomic Features*

Before extracting radiomic features from CT images, HU value range of CT images were normalized at first. In other words, Hu value of pixel in CT images larger than 1000 was set as 1000, and then send the images to extract features.

The masks of the ROIs (tumors) are stored in DICOM format in NSCLC Radiogenomics whilst the segmentation of each nodule is stored in XML file in the LIDC-IDRI dataset. The 3D masks for corresponding ROIs (tumors or nodules) were reconstructed from their corresponding files. We used pyradiomics [35] (version 2.2.0) to calculate 103 radiomic features for further analysis. All features included in the analyses are listed in the Supplementary Table 3.

*2.4 Model development*

One of the main tasks in the seminal article on radiomics by Aerts *et al.* [1] is survival prediction. For pre-treatment prediction of survival at 4 years, we used least squares support vector machines (SVMs) [37] with Radial Basis Function (RBF) Kernel as our classifier. SVMs use regularization to prevent overfitting when the number of input variables is high [36]. The input variables for the classifier were age and the 103 radiomic features extracted from the tumor.

For lung cancer diagnosis, we used deep attention based multiple instance learning [26] as the classifier as shown in paper [24]. The main characteristic of this classifier is that it can classify groups of samples (e.g. issue a diagnosis based on a set of CT scans from a patient) and reveal the importance of each sample in determining the diagnosis. The architecture of the method is shown in Supplementary Figure 2. The inputs of the model are the radiomic features as and the clinical diagnosis (cancer or not) is the output.

*2.5 Experiments*

We applied the trained generative models to denoise real low dose CT images before extracting the radiomic features. Subsequently, we trained the classification models for survival prediction and lung cancer diagnosis using radiomic features and we compared their performance with that of models trained using radiomic features extracted from low dose CT images.

All denoising experiments for low dose CT images were executed on a Core i7 8565 U CPU with 8GB of RAM based on pre-trained generative models. Based on training specifications described in [27], generative models were trained 25, 50, 75 and 100 epochs. All four trained models were used for denoising. For internal validation, 40 trials of nested cross validation [38] of RBF kernel SVM were executed and the number of GroupKFold in each trial was set as 5 for survival predication validation. We adopted the minority oversampling strategy described in [39] for lung cancer diagnosis task to improve the model's performance due to our dataset being small and imbalanced.

We assessed the models' performance calculating their area under the receiver-operating characteristics curve (AUC), accuracy and recall (using a probability threshold of 0.5). Finally, we used Student's T-test to assess the statistical significance of the differences in model performance results.

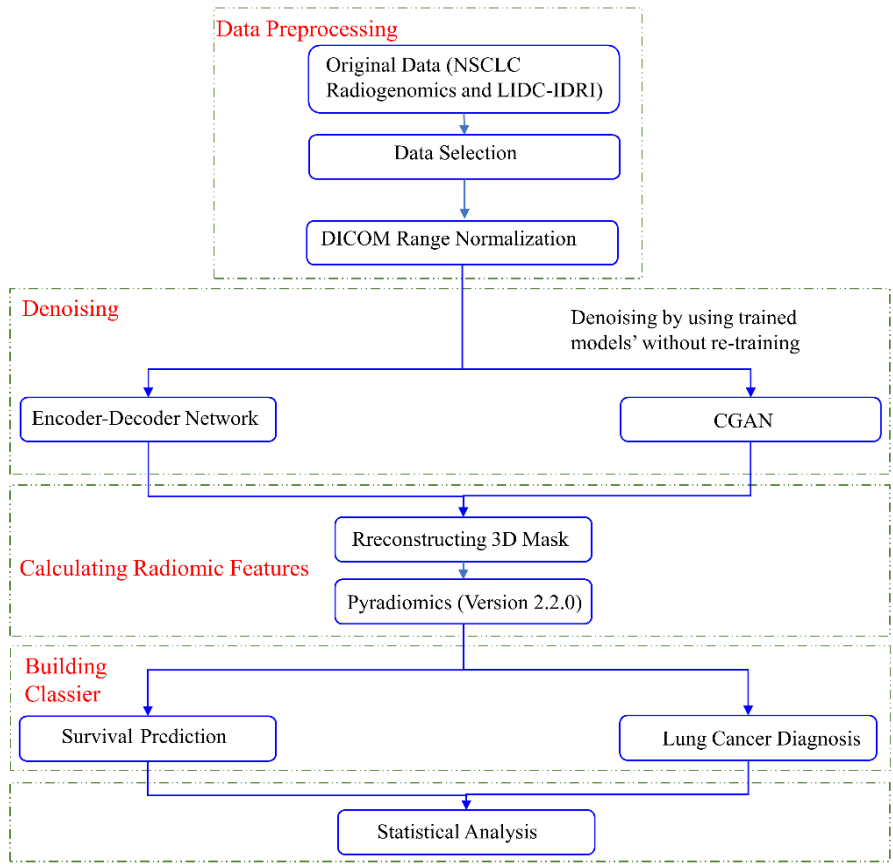

Figure 1. Flowchart of methods

**3. Results**

An example of an original CT image from the NSCLC Radiogenomics dataset and its denoised counterparts are shown in Figure 2.

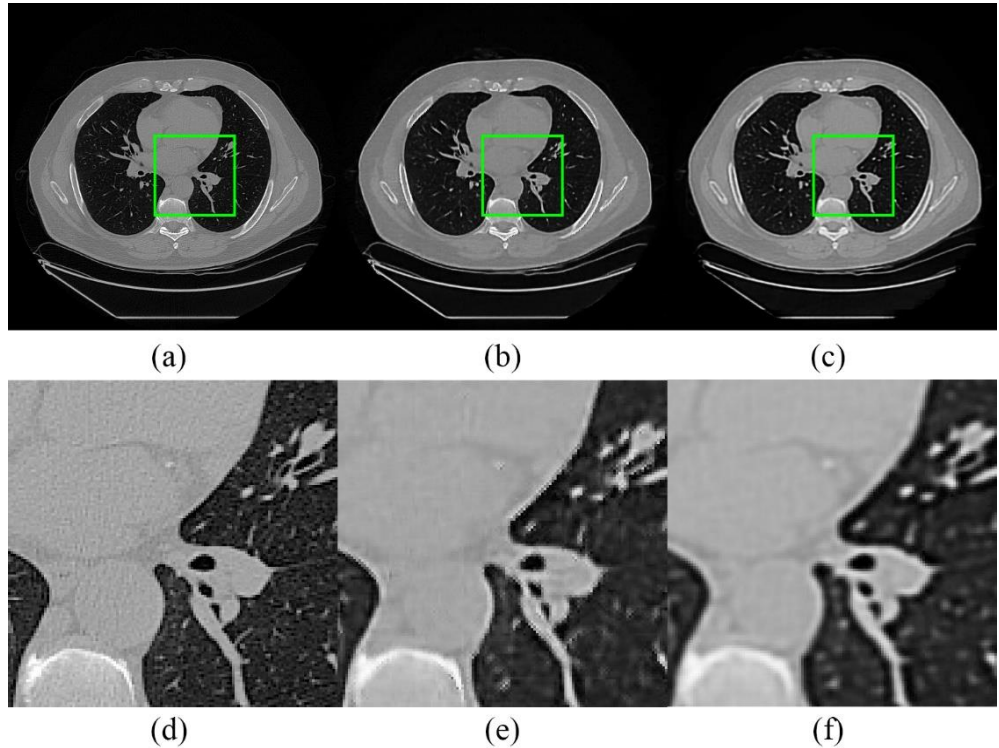

Figure 2. Example of low dose CT denoising: (a) original CT Image from NSCLC Radiogenomics (R01-003, radiation exposure: 7 mAs); (b) image denoised by the CGAN (100 epochs); (c) image denoised by the encoder-decoder network (100 epochs); (d) zoomed region of interests (ROI) of (a); (e) zoomed ROI of (b); and (f) zoomed ROI of (c)

*3.1 Survival Prediction*

The 4-year survival prediction model based on radiomic features extracted from low dose CTs achieved an AUC of 0.524 with a standard error of the mean (SEM) of 0.042. On the other hand, the survival prediction models based on radiomic features extracted from denoised low dose CTs achieved AUC ranging between 0.54 and 0.58. As shown in Table 1 and Figure 3, encoder-decoder networks and CGANs can improve radiomics-based models' performance significantly. The difference between encoder-decoder network and CGAN was not significant when trained for 75 epochs and 100 epochs, similar to what was reported in reference [27].

Table 1. Experimental results for 4-year survival prediction

| Training length / Metrics | Without Denoising | 25 Epochs | 50 Epochs | 75 Epochs | 100 Epochs |
|---|---|---|---|---|---|
| Encoder-decoder network | | | | | |
| AUC | 0.525 ± 0.042 | 0.580 ± 0.049 | 0.572 ± 0.040 | 0.554 ± 0.051 | 0.566 ± 0.044 |
| p-value * | -- | < 0.01 | < 0.01 | < 0.01 | < 0.01 |
| CGAN | | | | | |
| AUC | -- | 0.537 ± 0.045 | 0.551 ± 0.049 | 0.538 ± 0.123 | 0.566 ± 0.53 |
| p-value | -- | 0.20 | 0.01 | 0.16 | < 0.01 |
| Encoder-decoder network versus CGAN | | | | | |
| p-value ** | -- | < 0.01 | 0.04 | 0.15 | 0.93 |

*compared with results from original radiomics; ** comparing encoder-decoder network and CGAN.

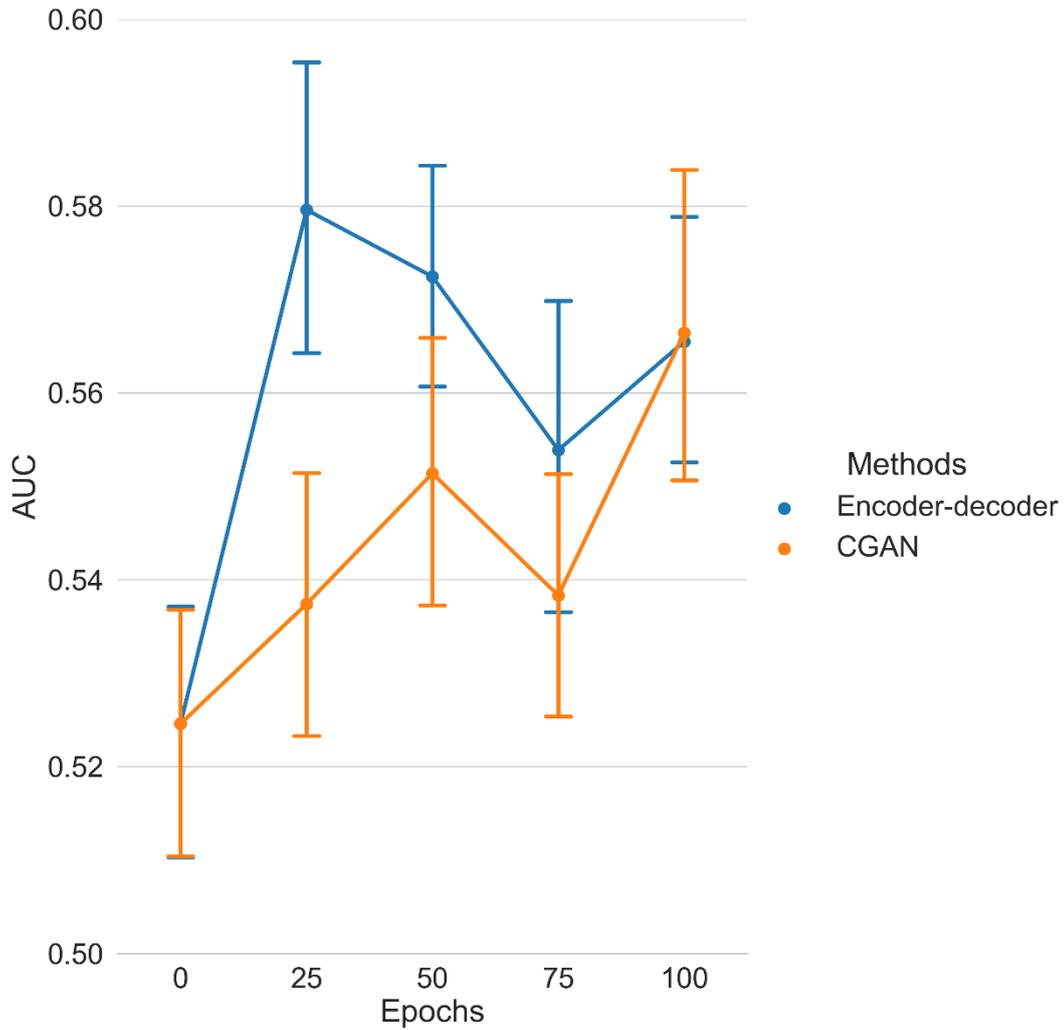

Figure 3. Experimental results (AUC) of survival prediction task

*3.2 Lung Cancer Diagnosis*

As shown in [24], our method can achieve an AUC of 0.842 (SEM 0.074) based on radiomic features extracted from the original low dose CT scans from the LIDC-IDRI dataset. The AUCs of the classification models based on radiomics extracted from denoised images range between 0.84 and 0.89 as shown in Table 2 and Figure 4(c). Models built using radiomic features calculated from denoised images outperformed models developed from the original radiomic features in most experiments. Similarly to survival prediction, the difference between encoder-decoder network and CGAN was not significant when trained for 75 and 100 epochs.

Table 2. The AUCs of different models for lung cancer diagnosis

| Training length<br>Metrics | Without Denoising | 25 Epochs | 50 Epochs | 75 Epochs | 100 Epochs |
|---|---|---|---|---|---|
| Encoder-decoder Network | | | | | |
| AUC | 0.842 ± 0.071 | 0.883 ± 0.077 | 0.844 ± 0.067 | 0.823 ± 0.067 | 0.866 ± 0.070 |
| p-value* | -- | < 0.01 | 0.86 | 0.07 | 0.02 |
| CGAN | | | | | |
| AUC | -- | 0.894 ± 0.056 | 0.863 ± 0.066 | 0.837 ± 0.085 | 0.866 ± 0.056 |
| p-value* | -- | < 0.01 | 0.06 | 0.49 | 0.01 |
| Differences of results by comparing Encoder-decoder network and CGAN | | | | | |
| p-value* | -- | 0.31 | 0.07 | 0.75 | 0.52 |

*compared with results from original radiomics;

Figure 4 (a) and (b) and Table 3 show that denoising had a negative impact in the accuracy and recall of the lung cancer diagnosis classification models, when using a threshold of 0.5.

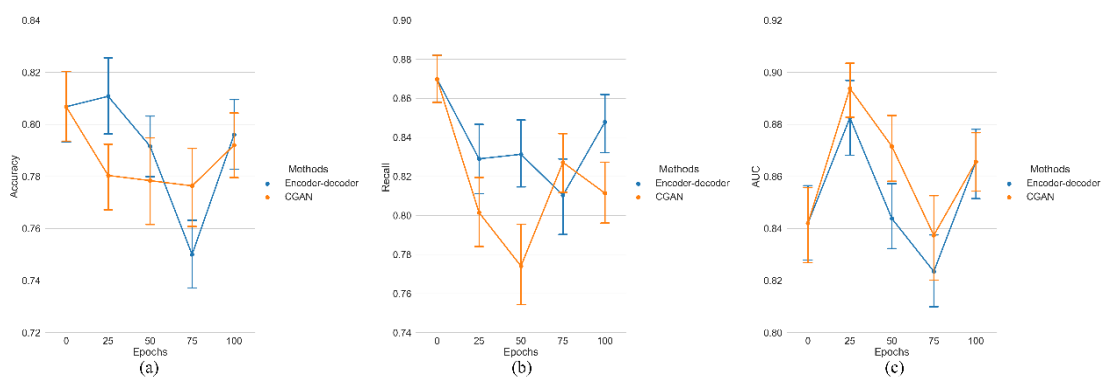

Figure 4. Experimental results of lung cancer diagnosis: (a) Accuracy, (b) recall and (c) AUC.

Table 3. Accuracy and recall for lung cancer diagnosis

| Training length<br>Metrics | 0 Epochs | 25 Epochs | 50 Epochs | 75 Epochs | 100 Epochs |
|---|---|---|---|---|---|
| Encoder-decoder network | | | | | |
| Accuracy | 0.807 ± 0.069 | 0.818 ± 0.077 | 0.792 ± 0.061 | 0.75 ± 0.067 | 0.796 ± 0.067 |
| p-value* | -- | 0.70 | 0.10 | < 0.01 | 0.26 |
| Recall | 0.870 ± 0.061 | 0.829 ± 0.096 | 0.831 ± 0.089 | 0.810 ± 0.097 | 0.848 ± 0.072 |
| p-value* | -- | < 0.01 | < 0.01 | < 0.01 | 0.02 |
| CGAN |   |   |   |   |   |
| Accuracy | -- | 0.780 ± 0.068 | 0.779 ± 0.085 | 0.776 ± 0.074 | 0.798 ± 0.064 |
| p-value* | -- | 0.01 | 0.01 | < 0.01 | 0.12 |
| Recall | -- | 0.802 ± 0.091 | 0.774 ± 0.104 | 0.827 ± 0.079 | 0.811 ± 0.079 |
| p-value* | -- | < 0.01 | < 0.01 | < 0.01 | < 0.01 |
| Encoder-decoder network versus CGAN (p-values) |   |   |   |   |   |
| Accuracy | -- | < 0.01 | 0.21 | 0.01 | 0.67 |
| Recall | -- | 0.04 | < 0.01 | 0.18 | < 0.01 |

*compared with results from original radiomics;

## 4. Discussion

In this study, we aimed to assess the potential of generative models to improve the performance of prediction models based on radiomic features extracted from low dose CT scans. The results show that encoder-decoder networks and CGANs can improve the AUC of radiomics for survival prediction and lung cancer diagnosis based on different low dose CT datasets. These findings imply that denoising low dose CT scans using generative models is a convenient pre-processing step before calculating radiomic features to train a predictive or diagnostic model.

The results also show that denoising using generative models might lead to a decrease in accuracy and recall. This might be caused by a shift in the receiver operating characteristic (ROC) curve as a result of the denoising. However, a higher AUC implies that there are other thresholds for which the accuracy and recall are higher with the denoised images. The threshold will differ for each possible application of these models, and a model with a higher AUC will be more likely to have a better accuracy/recall combination.

Another interesting aspect of the results is the variability of the models' AUCs for different numbers of training epochs. As shown in Figure 3 and Figure 4 (c), the performance of the models improves after the first epochs, then deteriorates when training for a higher number of epochs, and finally it seems to improve again after a particular number of training epochs. This tendency seems more significant in Figure 4 (c) than Figure 3. This might be explained by a phenomenon that has attracted considerable attention in the deep learning research domain in last few years -- deep double descent [40][41]. Unfortunately, the mechanisms of this phenomenon are still unclear, and more research on this topic is needed.

It is worth delving into the cause for the observed improvement using generative models. As mentioned previously, we think this improvement is brought on by the denoising effect of generative models to low dose CT. However, as shown in Supplementary Figure 1 (b), 40% CT images in the LIDC-IDRI dataset were not noisy (since they were scanned with over 200 mAs). Denoising these images using generative models would decrease images' quality. Therefore, there must be another source of improvement. Our hypothesis for this alternative source of improvement is dose normalization. In other words, generative models not only improved image quality of low dose CT images in dataset but also transfer the imaging exposure of the whole dataset from a wide range to a more compact but unknown range.

One potential limitation of our study is the low AUCs achieved by the models for pre-treatment survival prediction for lung cancer based on radiomic features. However, these are in line with results reported elsewhere. For example, Isensee *et al.* [42] reported an accuracy of 52.6% based on the BraTS 2017 dataset [43] for brain tumor by using radiomics; Choi *et al.* [44] reported an integrated AUC (iAUC) of 0.620 [95% CI: 0.501–0.756] in TCGA/TCIA dataset using random survival forest to derive a prediction model; Finally, Bae *et al.* [45] reported an iAUC of 0.590 [95% CI: 0.502, 0.689] for overall survival prediction in Glioblastoma using MRI radiomic features.

Regarding future work, we believe generative models should be trained to keep more information from the original domain. More specific, low level domain adaptation such as denoising for medical images should focus on keeping content information from original domain in the target domain. For example, by adding a content loss term in the cost function, adjusting generative models training method as shown in [47]. Second, more generative models with different architectures should be considered as the test models to find better models for this task. Finally, since the validity of the results of this study are limited to our selected datasets and tasks, further application to more datasets and tasks could reinforce or disprove our findings.

**5. Conclusion**

In this study, we assessed the potential of generative models (CGANs and encoder-decoder networks) to improve the performance of low dose CT scan radiomics-based models in two tasks – survival prediction and lung cancer diagnosis – and two datasets – NSCLC Radiogenomics and LIDC-IDRI. SVM and deep attention based multiple instance learning were used classifiers in survival prediction and lung cancer diagnosis respectively. The results support the hypothesis that generative models can improve radiomics performance in different tasks and datasets. In conclusion, denoising using generative models is an effective pre-processing step for calculating radiomic features from low dose CT.

**Declaration of Competing Interest**

The authors declare that they have no known competing financial interests or personal relationships that could have appeared to influence the work reported in this paper.

**Acknowledgments**

JC is supported by a China Scholarship Council scholarship (201906540036). The remaining authors acknowledge funding support from the following: STRaTegy (STW 14930), BIONIC (NWO 629.002.205), TRAIN (NWO 629.002.212), CARRIER (NWO 628.011.212) and a personal research grant by The Hanarth Funds Foundation for LW.